\documentclass[a4paper]{article}
\usepackage[backref]{hyperref}
\usepackage{INTERSPEECH2021}
\usepackage{cite}

\title{Speech Enhancement using Separable Polling Attention and Global Layer Normalization followed with PReLU}
\name{Dengfeng Ke$^1$, Jinsong Zhang$^1$, Yanlu Xie$^1$, Yanyan Xu$^2$, Binghuai Lin$^3$}
\address{
  $^1$Beijing Language and Culture University\\
  $^2$Beijing Forestry University\\
  $^3$Smart Platform Product Department,Tencent Technology Co., Ltd, China}
\email{\{dengfeng.ke,jinsong.zhang,xieyanlu\}@blcu.edu.cn, xuyanyan@bjfu.edu.cn, binghuailin@tencent.com}

\begin{document}

\maketitle
\begin{abstract}
Single channel speech enhancement is a challenging task in speech community.
Recently, various neural networks based methods have been applied to speech enhancement.
Among these models, PHASEN and T-GSA achieve state-of-the-art performances on the publicly opened VoiceBank+DEMAND corpus.
Both of the models reach the COVL score of 3.62. PHASEN achieves the highest CSIG score of 4.21 while T-GSA gets the highest PESQ score of 3.06.
However, both of these two models are very large. The contradiction between the model performance and the model size is hard to reconcile.
In this paper, we introduce three kinds of techniques to shrink the PHASEN model and improve the performance.
Firstly, seperable polling attention is proposed to replace the frequency transformation blocks in PHASEN.
Secondly, global layer normalization followed with PReLU is used to replace batch normalization followed with ReLU.
Finally, BLSTM in PHASEN is replaced with Conv2d operation and the phase stream is simplified.
With all these modifications, the size of the PHASEN model is shrunk from 33M parameters to 5M parameters,
while the performance on VoiceBank+DEMAND is improved to the CSIG score of 4.30, the PESQ score of 3.07 and the COVL score of 3.73.
\end{abstract}

\section{Introduction}

Single channel speech enhancement is one of the most challenging tasks in speech community.
Recently, due to the advances in deep learning, 
a great number of approaches for speech enhancement have been proposed, 
including various kinds of GAN-based models, 
such as SEGAN\cite{pascual2017segan},cnnGAN\cite{shah2018time},
 mmseGAN\cite{soni2018time}, serGAN\cite{baby2019sergan}, MetricGAN\cite{fu2019metricgan}, 
AeGAN\cite{abdulatif2020aegan}, ISEGAN and DSEGAN \cite{phan2020improving},
and various kinds of GAN-free models,
such as Wavenet\cite{reth2018wavenet}, Unet\cite{macartney2018unet}, 
AttUnet\cite{giri2019unet}, MDPhD\cite{kim2018multi}, DFL\cite{germain2019speech} and TCN\cite{koyama2020tcn}.

To the best of our knowledge, among all of the single channel speech enhancement systems,
PHASEN\cite{yin2020phasen}, T-GSA\cite{kim2020t}, NAAGN\cite{deng2020naagn} and
the phone-fortified-loss system\cite{hsieh2020improving} represent
the highest level of single channel speech enhancement on the VoiceBank+DEMAND corpus.

However, such systems are usually large in size and time consuming.
In practical use, we usually hope that the model is small and has good performance.
It is a very difficult problem to consider both the model size and the performances.

In this paper, we choose PHASEN as our baseline model,
and introduce the Separable Polling Attention to reduce the model size.
Moreover, Global Layer Normalization is found to converge faster than Batch Normalization,
and PReLU has more general adaptability than ReLU for both of the amplitude stream and the phase stream.
We can conclude that our proposed model has smaller size with better performances in many indices.

\section{PHASEN}

This section reviews the PHASEN\cite{yin2020phasen} system and its important components we want to redesign later.

\subsection{PHASEN Overview}
\begin{figure}[htbp]
  \centering
  \includegraphics[width=\linewidth]{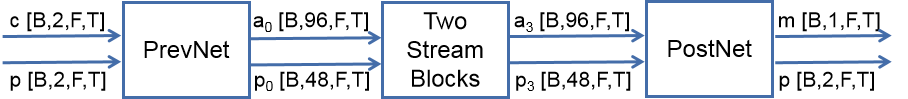}
  \caption{PHASEN Overview}
  \label{fig:PHASEN_overview}
  \vspace{-0.5cm}
\end{figure}

  As is shown in figure \ref{fig:PHASEN_overview},
  PHASEN receives two input streams, $c$ and $p$,
  which are the complex spectrogram and the phase-only spectrogram, respectively.
  The PrevNet projects $c$ from 2 channels to 96 channels,
  and $p$ from 2 channels to 48 channels, by $Conv2d$ operations.
  The Two Stream Blocks (TSB) process the amplitude stream $a_i$ and the phase stream $p_i$ parallelly
  and exchange information between $a_i$ and $p_i$ at the end of each block.
  Finally, the PostNet projects the amplitude stream $a_3$ down to a single-channel amplitude mask $m$
  and projects the phase stream $p_3$ down to a 2-channels phase estimation $p$.
  The mask $m$ is multiplied to the noisy amplitude spectrogram $a_{noisy}$ to fetch the estimated enhanced amplitude spectrogram $a$,
  and then $p$ is multiplied to $a$ and then inverse short time Fourier transform is used to get the enhanced speech signal.

\subsection{Two Stream Block}
\begin{figure}[htbp]
  \begin{center}
  \includegraphics[width=\linewidth]{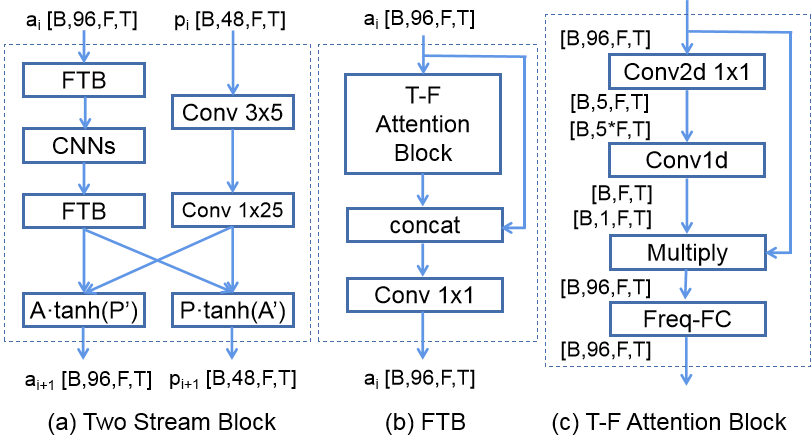}
  \caption{Two Stream Block}
  \label{fig:TSB_overview}
  \end{center}
  \vspace{-0.5cm}
\end{figure}

Two Stream Block (TSB), as shown in figure \ref{fig:TSB_overview}, is the key contribution of PHASEN.
In \cite{yin2020phasen}, Yin et al. pointed out that simply stacking several 2D convolutional layers with small kernels cannot capture global correlation among harmonics,
so they design Frequency Transformation Block (FTB) to capture full-frequency receptive field.
The main idea of FTB is to narrow down the channel number of the amplitude stream,
and then mixing up all channels and all frequencies with a 1D convolution,
so each frequency bin has an attention weight that considering all harmonics together with all channels.
The authors called this the Time-Frequency (T-F) Attention, which improves all the performance indices greatly.
However, this structure also results in almost 3 millions of parameters in each FTB
and 18 millions of parameters in the whole model.
That is why we plan to shrink it down.

\subsection{PostNet with BLSTM}
\begin{figure}[htbp]
  \begin{center}
  \includegraphics[width=\linewidth]{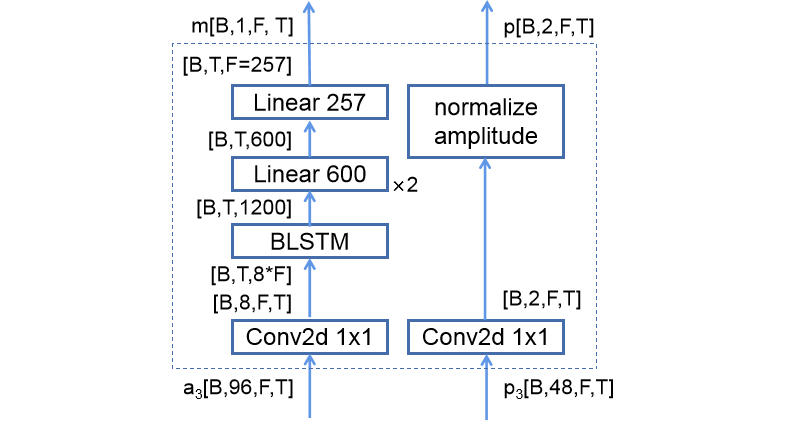}
  \caption{PostNet with BLSTM}
  \label{fig:postnet_overview}
  \end{center}
  \vspace{-0.5cm}
\end{figure}

As is shown in figure \ref{fig:postnet_overview},
the PostNet projects the amplitude stream and the phase stream back to short time Fourier transform domain.
The amplitude stream is narrowed down to 8 channels at first,
and then fed to BLSTM and three fully connected layers to predict the amplitude mask.
As the BLSTM maps 2,056 dimensional features to 600 dimensional space,
this leads to an extra 12 millions of parameters in the model.

\section{The Proposed Method}

\subsection{Separable Polling Attention}
In 2017, Collet \cite{chollet2017xception} introduced a \emph{depthwise separable convolution} operation,
namely a depthwise convolution followed by a pointwise convolution,
which attempts to learn filters in a 3D space with two spatial dimensions (width and height) and a channel dimension separately.
This separable architecture obtains small gains in accuracy and large gains in convergence speed, as well as significant reduction in model size.
There is another separable architecture named \emph{spatially separable convolution} that factors an n$\times$m convolution kernel into an n$\times$1 and a 1$\times$m kernels.
Mamalet et al. \cite{mamalet2012simplifying} reported that this spatially separable convolution speeds up learning and processing time with nearly the same level of recognition performances as classical ConvNets.

\begin{figure}[htbp]
  \begin{center}
  \includegraphics[width=\linewidth]{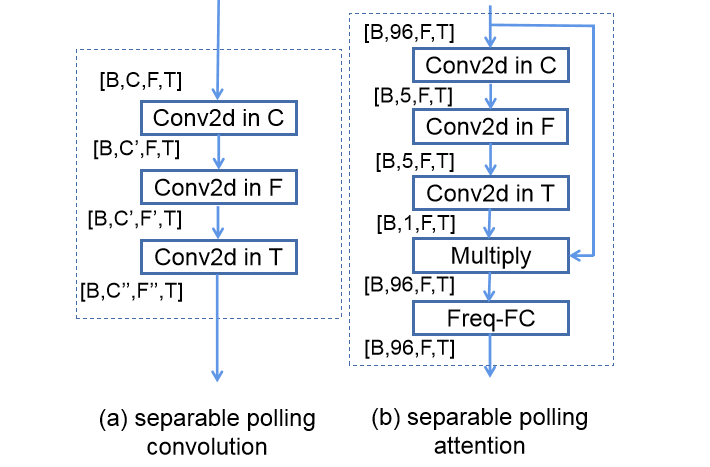}
  \caption{Separable Polling Attention}
  \label{fig:spa}
  \end{center}
  \vspace{-0.5cm}
\end{figure}

Inspired by these prior works, we proposed another kind of separable convolution suitable for speech.
As is known in the community, short time Fourier transform can be treated as a 1D convolution with Fourier bases,
and the outputs of this 1D convolution is usually reshaped to 2$\times$F$\times$T,
that is two channels to represent real and image parts of Fourier coefficients for each frequency bin for T frames.
Our separable polling convolution, in figure \ref{fig:spa}, works with a polling mechanism on this C$\times$F$\times$T shaped features of speech,
firstly a $1\times1$ convolution in channel dimension,
and sencondly a $1\times1$ convolution in frequency dimension
and finally a convolution conduced in time dimension.
In separable polling attention, the final convolution maps from C'=5 channels to C"=1 channel with a $1\times9$ kernel.
This separable polling attention architecture greatly shrinks down the size of the model with a slight improvement in speech enhancement.
In our experiments, the original T-F attention block is shrunk down from 2.8M parameters to 60K parameters.

Our separable polling convolution in figure \ref{fig:spa} has the following differences with depthwise separable convolution:
\begin{itemize}
\item Depthwise separable convolution usually performs first a channel-wise spatial convolution and then a $1\times1$ convolution in channel dimension.
      Our separable polling convolution performs first a $1\times1$ convolution in channel dimension
      and then a $1\times1$ convolution in frequency dimension (transpose of channel and frequency dimension is needed)
      and finally a common convolution in time dimension.
\item Each convolution in our separable polling convolution is followed with a global layer normalization (GLN) suggested in Conv-TasNet \cite{luo2019conv}
      and a parametric rectified linear unit (PReLU) proposed in \cite{he2015delving},
      while depthwise separable convolutions are usually implemented without non-linearities between separative operations.
\end{itemize}

\subsection{Global Layer Normalization with PReLU activation}

In 2015, Ioffe et al. \cite{ioffe2015batch} pointed out an \emph{internal covariate shift} phenomenon during network training.
They found that the distribution of each layer's inputs changes during training as the parameters of the previous layers change,
which requires the need for lower learning rates and carefully selected parameter initialization for the network.
Inspired by LeCun et al.'s finding \cite{lecun2012efficient} that the network training converges faster if its inputs are whitened,
batch normalization \cite{ioffe2015batch} is proposed to apply to each mini-batch to ensure the distribution invariance for each layer,
which also allows us to use much higher learning rates and be less careful about initialization.

However, the performance of batch normalization is greatly dependent on the mini-batch size.
Furthermore, it is not obvious how to apply it to recurrent neural networks.
Jimmy et al. \cite{ba2016layer} transpose batch normalization into layer normalization
by computing the statistics used for normalization from the neurons in a layer on a single training case.
They found that layer normalization works better than batch normalization especially
when the batch size is as small as 4.

\begin{figure}[htbp]
  \centering
  \includegraphics[width=\linewidth]{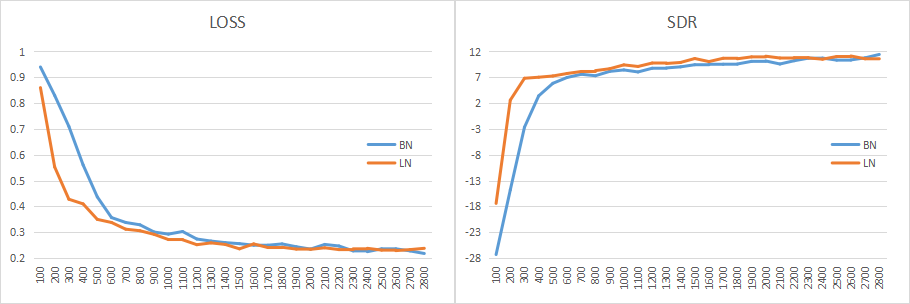}
  \caption{Batch Normalization vs. Layer Normalization}
  \label{fig:normalization}
  \vspace{-0.5cm}
\end{figure}

Yin et al.'s study \cite{yin2020phasen} reported that a performance drop of 0.97dB on SDR and a performance drop of 0.12 on PESQ
are observed if global layer normalization is used in amplitude stream.
However, in our study, global layer normalization always outperforms batch normalization
within the first 1,000 steps as is shown in figure \ref{fig:normalization},
but as the number of iterations increases, they show no significant difference.
This maybe come from the fact that we carry on our study on GTX1080Ti with only 11G memory,
 leading to smaller batch size than Yin et al.'s.
Moreover, we find that with this smaller batch size,
our re-implementation of PHASEN performs slightly better than that Yin et al. proposed in \cite{yin2020phasen}.

\begin{figure}[htbp]
  \begin{center}
  \includegraphics[width=\linewidth]{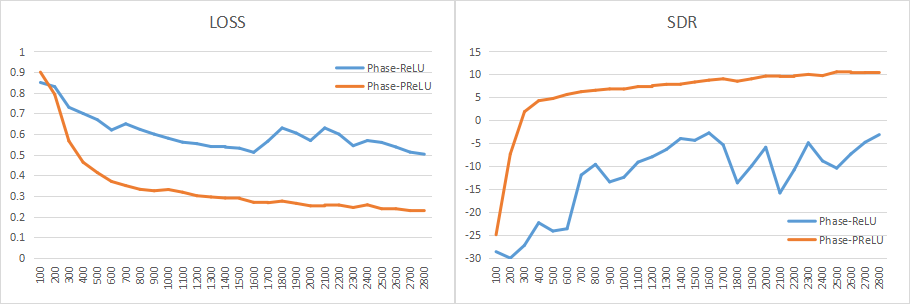}
  \caption{ReLU vs. PReLU for Phase Stream}
  \label{fig:phase-relu-prelu}
  \includegraphics[width=\linewidth]{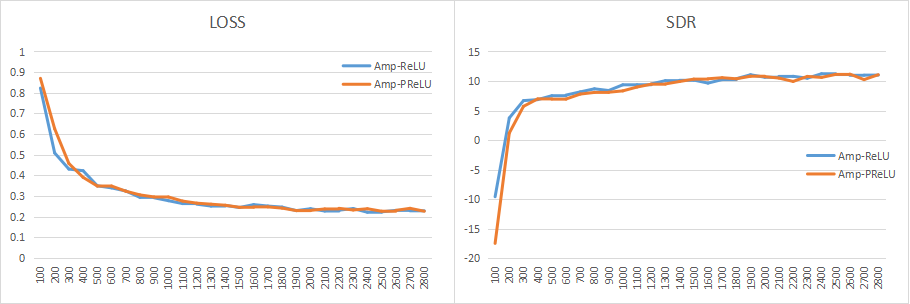}
  \caption{ReLU vs. PReLU for Amplitude Stream}
  \label{fig:amp-relu-prelu}
  \end{center}
  \vspace{-0.5cm}
\end{figure}

In \cite{yin2020phasen}, Yin et al. also suggest to use the rectified linear unit (ReLU) activation function \cite{nair2010rectified}
 on the amplitude stream but no activation function on the phase stream.
Our experiments also support this founding that the performance degrades dramatically if ReLU is applied to the phase stream, as illustrated in figure \ref{fig:phase-relu-prelu}.
However, we also find that if the Parametric Rectified Linear Unit (PReLU) activation \cite{he2015delving} is applied to the phase stream,
the performance degradation disappears, as illustrated in figure \ref{fig:phase-relu-prelu}.
When we apply ReLU and PReLU to the amplitude stream they show no significant difference, as illustrated in figure \ref{fig:amp-relu-prelu}.

Considering the above phenomena, we employ the Global Layer Normalization (GLN) used in Conv-TasNet \cite{luo2019conv}
followed with Parametric Rectified Linear Unit (PReLU) activation suggested in \cite{he2015delving} between convolutional layers and linear layers.

\subsection{Simplifying PostNet and Phase Stream}

\begin{figure}[htbp]
  \begin{center}
  \includegraphics[width=\linewidth]{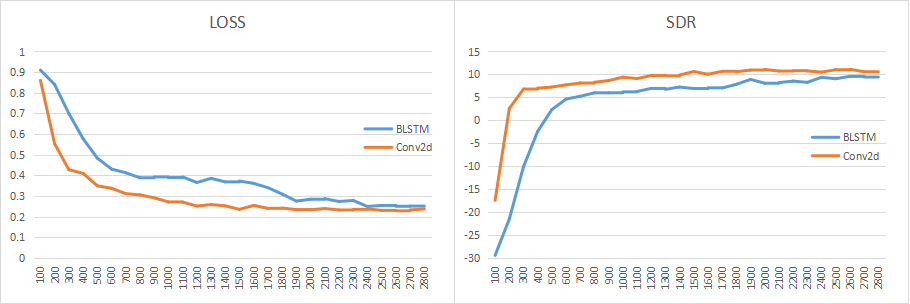}
  \caption{BLSTM vs. Conv2d}
  \label{fig:blstm-conv2d}
  \end{center}
  \vspace{-0.5cm}
\end{figure}

As mentioned above, BLSTM is the biggest layer, about 12 millions of parameters, in PHASEN.
We find that, this layer not only takes up memory but also slows down convergency.
As is shown in figure \ref{fig:blstm-conv2d},
when the BLSTM layer is replaced with Conv2d operation with 600 filters, down to about 1 million of parameters,
the loss decreases rapidly while the SDR increases rapidly.
We also find that, directly transforming the amplitude stream from 96 channels to 2 channels in PostNet,
or shrinking the model with dilated convolution leads to performance degradation.
Moreover, the phase stream is important to help the amplitude stream,
but simplifying the phase stream will not affect performances,
so we use a single Conv2d operation with the kernel size of 3 in frequency dimension and 5 in time dimension for the phase stream in each TSB block.

\begin{table*}[htbp]
  \caption{Performance comparison on VoiceBank+DEMAND}
  \label{tab:performance-comparison-on-VoiceBank+DEMAND}
  \begin{center}
  \begin{tabular}{l|c|c|c|c|c|c|c}
    \toprule
    \textbf{Method}                     &\textbf{YEAR} &\textbf{SSNR}  &\textbf{PESQ} &\textbf{CSIG} &\textbf{CBAK} &\textbf{COVL} &\textbf{SDR} \\
    \midrule
    Noisy                               &-             &1.68           &1.97          &3.35          &2.44          &2.63          &8.44         \\
    \midrule
    SEGAN\cite{pascual2017segan}        &2017          &7.73           &2.16          &3.48          &2.94          &2.80          &-            \\
    Unet\cite{macartney2018unet}        &2018          &9.97           &2.40          &3.52          &3.24          &2.96          &-            \\
    WaveNet\cite{reth2018wavenet}       &2018          &-              &-             &3.62          &3.23          &2.98          &-            \\
    mmseGAN\cite{soni2018time}          &2018          &-              &2.53          &3.80          &3.12          &3.14          &-            \\
    MDPhD\cite{kim2018multi}            &2018          &10.22          &2.70          &3.85          &3.39          &3.27          &-            \\
    DFL\cite{germain2019speech}         &2018          &-              &-             &3.86          &3.33          &3.22          &-            \\
    AttUnet\cite{giri2019unet}          &2019          &10.05          &2.57          &3.79          &3.32          &3.18          &-            \\
    TCN\cite{koyama2020tcn}             &2020          &-              &2.89          &4.24          &3.40          &3.55          &-            \\
    PHASEN\cite{yin2020phasen}          &2020          &10.18          &2.99          &4.21          &3.55          &3.62          &-            \\
    NAAGN\cite{deng2020naagn}           &2020          &10.25          &2.90          &4.13          &3.50          &3.51          &-            \\
  T-GSA\cite{kim2020t}(enc:50M,dec:60M) &2020          &\textbf{10.78} &3.06          &4.18          &3.59          &3.62          &19.57        \\
    \midrule
    PHASEN-1(33M, our imp.)             &2021          &10.36          &3.03          &4.26          &3.57          &3.65          &19.44        \\
    PHASEN-2(33M, our imp.)             &2021          &10.42          &3.01          &4.22          &3.56          &3.61          &19.49        \\
    \midrule
    SPA-BN-PReLU-1(5M, our proposed)    &2021          &10.37          &3.02          &4.27          &3.59          &3.66          &19.46         \\
    SPA-BN-PReLU-2(5M, our proposed)    &2021          &10.42          &3.05          &4.30          &\textbf{3.61} &3.69          &19.55         \\
    \midrule
    SPA-LN-ReLU-1(5M, our proposed)     &2021          &10.11          &\textbf{3.07} &\textbf{4.30} &3.60          &\textbf{3.73} &19.24   \\
    SPA-LN-ReLU-2(5M, our proposed)     &2021          &10.37          &3.04          &4.29          &\textbf{3.61} &3.70          &19.47   \\
    \midrule
    SPA-LN-PReLU-1(5M, our proposed)    &2021          &10.55          &3.03          &4.28          &3.60          &3.67          &\textbf{19.62} \\
    SPA-LN-PReLU-2(5M, our proposed)    &2021          &10.49          &\textbf{3.07} &\textbf{4.30} &\textbf{3.61} &\textbf{3.73} &19.58         \\
    \bottomrule
  \end{tabular}
  \end{center}
  \vspace{-0.5cm}
\end{table*}

\section{Experiments and Results}

\subsection{Experimental Settings}
\textbf{Dataset}:
All our experiments are evaluated on the open dataset VoiceBank+DEMAND \footnote{https://datashare.ed.ac.uk/handle/10283/1942}
widely used for speech enhancement proposed in \cite{valentini2016speech}.
The corpus consists of 30 speakers from the Voice Bank corpus \cite{veaux2013voice}
and 8 noise conditions from the DEMAND corpus \cite{thiemann2013diverse} and 2 types of artificially generated noise.
The training set contains 11,572 utterances from 28 speakers with 5 types of noise,
while the test set contains 824 utterances from the other 2 speakers with the other 5 types of noise.

\textbf{Evaluation metrics}:
We use the same metrics with other papers, including
segmental signal-to-noise ratio (SSNR), perceptual evaluation of speech quality (PESQ),
composite signal distortion (CSIG), composite background intrusiveness (CBAK),
composite mean opinion score (COVL), and signal to distortion ratio (SDR).
All these metrics are open-sourced in this package  \footnote{https://www.crcpress.com/downloads/K14513/K14513\_CD\_Files.zip}
except that the SDR is implemented by ourselves according to formula (19) in \cite{kim2019end}.

\textbf{Parameters}:
Short time Fourier transform (STFT) is performed with the hanning window of size 512 and the shifting size of 160 (namely 10ms).
The channel number is 96 for the amplitude stream and 48 for the phase stream.
The separable polling attention maps 96 channels to 5 channels with a $1\times1$ kernel,
then a $1\times1$ convolution in frequency dimension from 257 frequency bins to 257 frequency bins,
and finally a time dimensional convolution with the kernel size of 9 to a single channel.

\textbf{Losses}:
We use the same loss function as is proposed in PHASEN \cite{yin2020phasen},
which sums up amplitude loss and phase-aware loss with equal weights.
The amplitude loss is the mean square error loss for amplitude with 0.3 power-law, and
the phase-aware loss is the mean square error loss computed from amplitude with 0.3 power-law in the estimated phase direction.

\textbf{Others}:
Adam optimizer is used to train each model with a warmup learning rate growing from 0.0000 to 0.0002 for the first 6000 steps without learning rate descending after reaching the peak.
Utterances are clipped to segments of 3 seconds randomly at each step with a mini-batch of 4 utterances. No data augmentation or dropout is used during training.
Each training uses only a single GTX1080Ti with 50 epoches.

\subsection{Comparing with Previous Systems}

Table \ref{tab:performance-comparison-on-VoiceBank+DEMAND} shows the performance comparison of the representative methods on VoiceBank+DEMAND corpus.
Representative results are listed according to the publishing time of the papers.

In order to check whether the performance of the model is stable, we have carried out 10 times of the experiments for each model, and
only two representative results are given in the table (denoted as X-1 and X-2). \textbf{our imp.} means our implementation of the model.
\textbf{SPA, -BN, -LN, -PReLU} and \textbf{-ReLU} represents our proposed Separable Polling Attention, with Batch Normalization, with Layer
Normalization, with PReLU and with ReLU on the amplitude stream, respectively.

Note that if ReLU is applied to the phase stream, all metrics will drop dramatically, namely,
PESQ from 3.07 to 2.15, CSIG from 4.30 to 3.63, CBAK from 3.61 to 2.58, COVL from 3.73 to 2.87.
The table only shows the results of ReLU applied to the amplitude stream.

Although our SPA model has only 5M parameters,
it works slightly better than PHASEN with 33M parameters.
Moreover, we find that our model converges much faster than PHASEN \cite{yin2020phasen}.

Our SPA-LN-PReLU model has the highest scores for all the performance metrics except the SSNR.
Although T-GSA has the highest SSNR score, its encoder contains 50M parameters according to\cite{kim2020t},
which is 10 times in size as ours. So our model is highly competitive in the case of the same size.

Experiments also find that layer normalization is slightly better than batch normalization,
and PReLU is more suitable for both the amplitude stream and the phase stream.
So global layer normalization followed with PReLU would be a better choice for our SPA model.

\section{Conclusions}

We proposed a separable polling attention model for speech enhancement,
which can reduce model size and accelerate convergence speed,
 and improve model performance at the same time. Our contribution includes:

\begin{enumerate}
  \item separable polling attention to greatly reduce the size with slight improvement of performances,
  \item layer normalization with parametric rectified linear unit to accelerate convergence speed,
  \item pruning unnecessary BLSTM in PostNet and unnecessary operations in the phase stream to accelerate training without performance degradation.
\end{enumerate}

\bibliographystyle{IEEEtran}
\bibliography{interspeech2021}

\end{document}